\documentclass[12pt]{article}
\usepackage[a4paper, margin=1in]{geometry}
\usepackage{authblk}
\usepackage{titlesec}
\usepackage{setspace}
\usepackage[numbers]{natbib}
\usepackage{indentfirst}
\usepackage{graphicx}
\usepackage{listings}
\usepackage{xcolor}
\usepackage{tabularx}
\usepackage{longtable}
\usepackage{array}
\usepackage{ragged2e}
\usepackage{booktabs}
\usepackage{makecell}
\usepackage{placeins}
\usepackage{xurl} % better line breaks for long URLs
\usepackage[
    colorlinks=true,
    linkcolor=blue,
    citecolor=blue,
    urlcolor=blue,
    breaklinks=true
]{hyperref}

\title{\textbf{A Multimodal Dataset for Large Language Model Applications in the Energy Domain}}

\author[1]{Costas Mylonas\thanks{\texttt{kmylonas@ubitech.eu}}}
\author[1]{Magda Foti\thanks{\texttt{mfoti@ubitech.eu}}}
\affil[1]{Energy Digitalization Group, UBITECH, Athens, Greece}

\date{} % No date

\begin{document}

\maketitle

\begin{abstract}

\noindent This paper presents the mAIEnergy dataset, an open-access, multimodal corpus developed to support Large Language Model (LLM) applications in the energy sector. The dataset integrates approximately 50,000 textual documents, 20,000 images, 25 million numerical time series records, and 2 million geospatial and relational data entries. It includes policy and regulatory texts, scientific articles and news articles, satellite and contextual imagery, electricity system measurements, weather observations, statistical indicators, and geospatial representations of energy infrastructure and related entities. All data have been harmonized into structured, ready-to-use formats, accompanied by consistent metadata and reproducible data retrieval and preparation workflows. The dataset can serve as a foundational energy knowledge base, allowing energy stakeholders to integrate additional open-source or proprietary data. The mAIEnergy dataset adheres to Findable, Accessible, Interoperable, and Reusable (FAIR) principles, enhancing its applicability for AI-driven energy research, modeling, and decision-making.

\end{abstract}

\section*{Background \& Summary}

The energy sector is undergoing rapid transformation driven by ambitious sustainability targets, decarbonization initiatives, and increasing digitalization. Effectively navigating these changes relies on advanced analytical tools capable of processing diverse and complex energy-related data. Artificial Intelligence (AI), particularly through Large Language Models (LLMs) and Retrieval-Augmented Generation (RAG), has shown significant potential for energy-focused tasks, including information retrieval, knowledge integration, and decision-support applications \cite{majumder2024exploring}. However, the effectiveness of these AI solutions is highly dependent on access to AI-ready, multimodal, and domain-specific datasets.

Currently, publicly available energy datasets typically focus on single modalities. For example, numerical data are provided from sources like the European Network of Transmission System Operators for Electricity (ENTSO-E) Transparency Platform \cite{hirth2018entso, ENTSOE} or Eurostat statistics \cite{Eurostat}. Geospatial data come from infrastructure mapping services, such as OpenStreetMap \cite{haklay2008openstreetmap}, while textual information is distributed across policy and research documents. Similarly, satellite imagery is sourced from platforms like the European Copernicus \cite{copernicus_discomap}. This fragmented landscape constrains their applicability for developing advanced AI-driven decision-support tools and conversational assistants, which rely on integrated multimodal datasets to achieve contextual understanding and reasoning capabilities \cite{lewis2020retrieval}.

Furthermore, the adoption of Open and Findable, Accessible, Interoperable, Reusable (FAIR) data practices, as defined by the relevant guiding principles \cite{wilkinson2016fair}, is becoming increasingly essential in the research community, including energy research. Nevertheless, the energy sector has been found to lag behind other domains in adopting comprehensive open-data and open-source practices \cite{pfenninger2017importance}. Bridging this gap requires the creation of harmonized, multimodal datasets that comply with FAIR principles and foster transparency, accessibility, and interoperability across the energy research ecosystem.

In response to this need, we introduce the mAIEnergy dataset \cite{mylonas2025maienergy}, an open-access corpus specifically designed to advance LLM-driven applications and research in the energy domain. The dataset integrates four distinct data modalities, namely textual, imagery, numerical, and geospatial to provide a comprehensive representation of the energy landscape.

Textual sources include articles from Wikipedia \cite{wikipedia}, news articles retrieved via the GNews API \cite{gnews}, scientific literature from arXiv \cite{arxiv}, and official European Union (EU) governmental documents. The latter comprise documents from sources, such as EU Directorate-General for Energy \cite{eu_dg_energy}, the Agency for Cooperation of Energy Regulators (ACER) \cite{acer} and national energy regulatory and ministries of EU countries. 

Imagery data sources comprise satellite images from Copernicus, which provides high-resolution observations of land use and energy infrastructure. Aerial images from the INRIA Aerial Image Labeling dataset \cite{inria_dataset} depict urban areas and support spatial and building analysis. Product-level energy labels from the European Product Registry for Energy Labelling (EPREL) \cite{eprel} represent standardized appliance efficiency classes, while façade imagery from the Irregular Facades (IRF) dataset \cite{irf_dataset} illustrates architectural and construction characteristics of buildings. In addition, open-license contextual images from Wikimedia Commons \cite{wikimedia_commons} and Wikipedia enrich the dataset with visual documentation of energy technologies, infrastructure, equipment, graphs, etc.

The numerical datasets include electricity load, generation, and market price data from ENTSO-E, together with annual energy balances and macroeconomic indicators from Eurostat. They also incorporate detailed building stock characteristics and performance metrics from the EU Building Stock Observatory (BSO) \cite{eu_building_stock}, as well as historical weather observations from Open-Meteo \cite{open_meteo}.

Finally, the geospatial and relational datasets were collected from OpenStreetMap, covering assets such as power plants, substations, renewable energy facilities, and Electric Vehicle (EV) charging infrastructure. Additional sources include the GridKit European Transmission Grid, which provides high-voltage network nodes and links \cite{gridkit}, the Global Power Plant Database, which details plant-level attributes \cite{wri_powerplants}, the ENTSO-E Transmission System Operator (TSO) Network, describing interconnections of the European transmission network, and the Community Research and Development Information Service (CORDIS) EU Projects Database, which maps EU-funded energy research projects and collaborations \cite{cordis}.

By integrating these modalities, the mAIEnergy dataset serves as an energy-domain knowledge base, enabling Continual Pre-Training (CPT) of LLMs and the development of RAG systems. The success of domain-specific knowledge bases in fostering specialized LLMs, such as Galactica for scientific reasoning \cite{taylor2022galactica}, demonstrates the value of curated, structured corpora for knowledge-intensive AI. Concurrently, the growing momentum of LLM research in the energy domain \cite{majumder2024exploring, antonesi2025transformers, jatowt2025flexidigital} emphasizes the need for an equivalent resource to advance energy-focused applications. Our multimodal approach enhances the effectiveness of LLM-powered conversational agents, enabling decision-support across various energy-related applications.

To position the mAIEnergy dataset within the existing landscape, Table~\ref{tab:dataset-comparison} compares it with representative public datasets along the dimensions of size, modalities, domain coverage, and accessibility. Existing energy datasets are predominantly single-modality. BuildingsBench \cite{emami2023buildingsbench} provides large-scale building load time series for short-term load forecasting, and Open Power System Data \cite{wiese2019open} compiles European load, generation, capacity, and price series together with power-plant tables for energy-system modelling. The energy-domain resources that target large language models remain text-only: EnergyGPT \cite{chebbi2026towards} is a corpus of scientific and general text used to specialise a general-purpose model for the energy sector, while ElecBench \cite{zhou2024elecbench} is a question-answering benchmark for evaluating models on power-dispatch tasks. Genuinely multimodal datasets exist, but in adjacent domains and are designed for supervised learning rather than for retrieval or generation. CropNet \cite{lin2024open} combines satellite imagery, gridded meteorological data, and crop statistics across United States counties for climate-aware yield prediction; METER-ML \cite{zhu2022meter} couples multi-sensor aerial and satellite imagery with geospatial annotations to map methane-emitting energy infrastructure; the Multimodal Flood Event Dataset \cite{zhang2024global} pairs multi-source remote-sensing imagery with heterogeneous text for global flood events; and SEN12MS \cite{schmitt2019sen12ms} provides co-registered radar and optical imagery for land-cover classification. Climate-focused text corpora such as ClimateBERT \cite{webersinke2021climatebert} support domain-adaptive language-model pre-training but contain only text.

\begin{table}[ht]
\centering
\scriptsize
\setlength{\tabcolsep}{4pt}
\renewcommand{\arraystretch}{1.25}
\caption{Comparison of the mAIEnergy dataset with representative public datasets in the energy, climate, and sustainability domains, along the dimensions of domain coverage, modalities, size, and accessibility.}
\label{tab:dataset-comparison}
\begin{tabularx}{\textwidth}{%
 >{\hsize=0.95\hsize\RaggedRight\arraybackslash\scriptsize}X
 >{\hsize=0.95\hsize\RaggedRight\arraybackslash\scriptsize}X
 >{\hsize=1.00\hsize\RaggedRight\arraybackslash\scriptsize}X
 >{\hsize=1.25\hsize\RaggedRight\arraybackslash\scriptsize}X
 >{\hsize=0.85\hsize\RaggedRight\arraybackslash\scriptsize}X}
\toprule
\textbf{Dataset} & \textbf{Domain coverage} & \textbf{Modalities (count)} & \textbf{Size / scale} & \textbf{Accessibility \& primary use} \\
\midrule
\textbf{mAIEnergy (this work)} & Energy (generation, grids, buildings, weather, regulation) & Textual, imagery, numerical, geospatial (4) & $\sim$50k documents; $\sim$20k images; $\sim$25M numerical records; $\sim$2M geospatial entries & Open, CC BY 4.0, Zenodo; LLM/RAG knowledge base \\
EnergyGPT \cite{chebbi2026towards} & Energy (general) & Textual (1) & $\sim$2.1B tokens of scientific and general text & Code open, corpus partly licensed; LLM specialisation \\
ElecBench \cite{zhou2024elecbench} & Electric power dispatch & Textual / QA (1) & 34,030 QA entries across 24 datasets & Open; LLM evaluation benchmark \\
BuildingsBench \cite{emami2023buildingsbench} & Building energy & Numerical time series (1) & 900k simulated buildings; evaluation suite of 7 real-building datasets & Open; load-forecasting ML \\
Open Power System Data \cite{wiese2019open} & European power systems & Numerical, tabular (1) & National and zonal load, generation, capacity, and price series; plant tables & Open (CC BY where permitted); energy-system modelling \\
METER-ML \cite{zhu2022meter} & Energy infrastructure / methane & Imagery, geospatial (2) & 86,599 georeferenced multi-sensor images & Open, CC BY 4.0; methane-source mapping ML \\
CropNet \cite{lin2024open} & Agriculture / climate & Imagery, meteorological, crop tabular (3) & Terabyte-scale; 2,291 US counties; 2017--2022 & Open; crop-yield ML \\
Multimodal Flood Event Dataset \cite{zhang2024global} & Flood / disaster & Textual, imagery (2) & Global flood events; 2000--2018 & Open, CC BY; flood-analysis ML \\
SEN12MS \cite{schmitt2019sen12ms} & Remote sensing / land cover & Imagery (radar + optical) (1) & 180,662 co-registered image triplets & Open; land-cover ML \\
ClimateBERT corpus \cite{webersinke2021climatebert} & Climate-related text & Textual (1) & $\sim$2.05M paragraphs (news, abstracts, reports) & Model open, corpus partly licensed; language-model pre-training \\
\bottomrule
\end{tabularx}
\end{table}

Two observations follow. First, to the best of our knowledge, no openly available corpus integrates textual, imagery, numerical, and geospatial modalities for the energy domain and is purpose-built for large language model and retrieval-augmented generation use, since the most modality-rich comparators span at most three modalities and target agriculture, land cover, or disaster response. Second, the contribution of the mAIEnergy dataset is not the co-location of these sources but the value added on top of them, which takes three concrete forms. The first is harmonisation and enrichment: the heterogeneous sources are converted into a single schema with consistent attribute names, normalised units, and per-record provenance metadata, and they are augmented with derived annotations that are not present in the raw sources, such as reverse-geocoded ISO country codes and explicit node and relationship tables that encode the relations between entities. The second is cross-modal linkage: records in different modalities are tied to a common set of energy-system entities through shared keys such as country, bidding zone, and geolocation, and a property graph connects these entities explicitly, so that the electricity-system series, the satellite tiles, the infrastructure assets, and the documents that refer to the same area can be retrieved together. The third is a semantic enrichment layer: the harmonised records are embedded and indexed for semantic, lexical, image, and graph-based retrieval, so that the dataset is directly usable for retrieval-augmented generation rather than requiring this layer to be built from scratch. Together with end-to-end coverage of the energy system from generation infrastructure and grids to buildings, weather, and regulation, and a single CC BY 4.0 release under a persistent identifier, these three layers distinguish the dataset from existing data-integration efforts, which typically aggregate sources within a single modality or domain without cross-modal linkage or retrieval-ready semantic representations.

Recognizing the dynamic nature of the energy domain, the mAIEnergy dataset is designed to be extensible, allowing users to utilize the corpus as-is, configure retrieval parameters to expand its scope, or integrate proprietary data to suit specific applications. To maximize usability and transparency, the dataset adheres to FAIR principles, providing structured metadata and openly sharing reproducible data retrieval and preparation workflows, together with the database back-ends used to index the corpus, via GitLab \cite{maienergy2024dataretrieval, maienergy2024vectordatabases}. This framework facilitates validation and enhancement by the broader community, ultimately aiming to accelerate the adoption of LLM-driven solutions and empower stakeholders to navigate the energy transition through informed, data-driven decision-making.

\section*{Methods}

\subsection*{Overview of the Workflow}

\begin{figure}[ht]
    \centering
    \includegraphics[width=\textwidth]{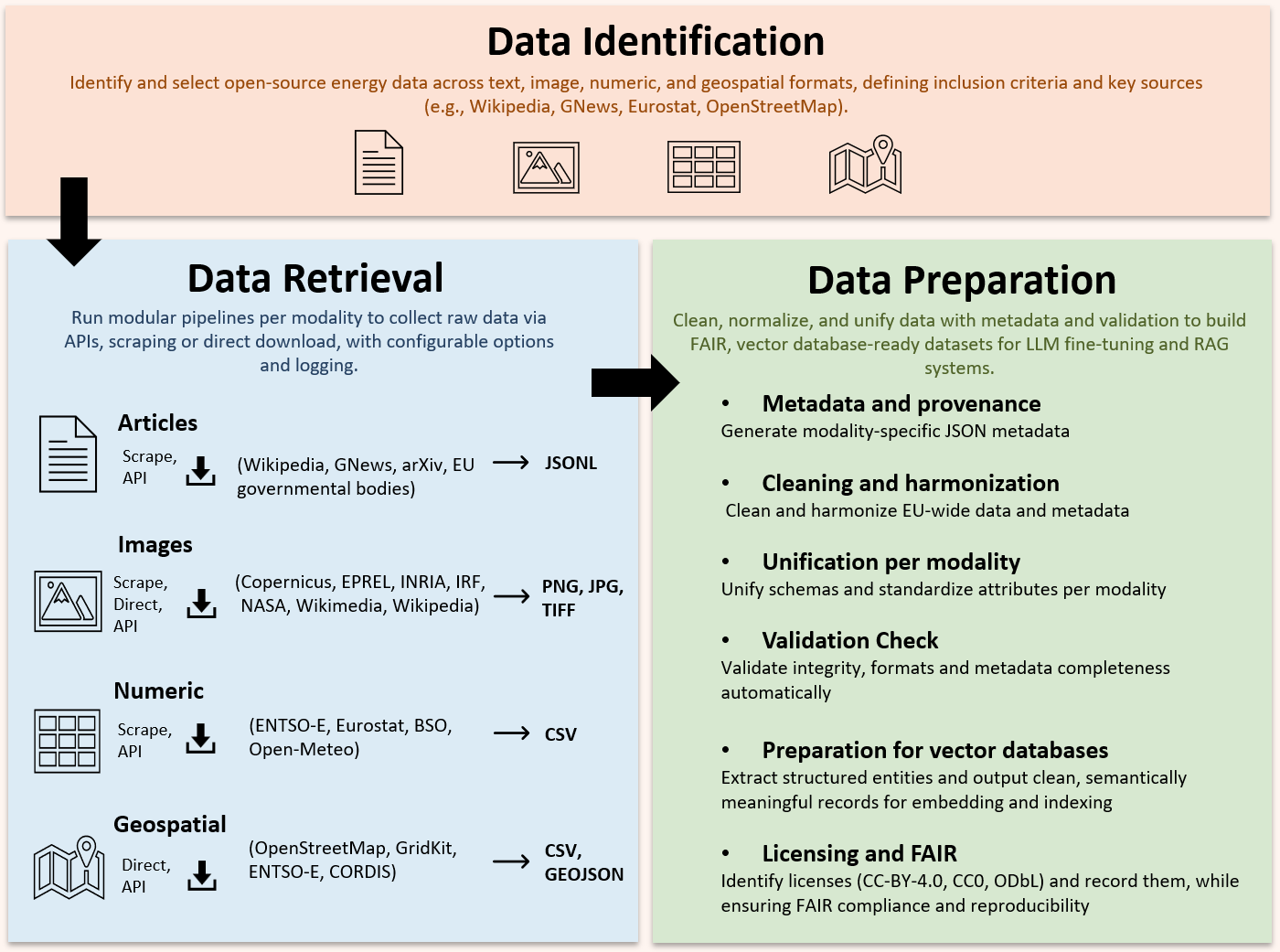} 
    \caption{Overview of the workflow used to generate the mAIEnergy dataset.}
    \label{fig:workflow}
\end{figure}

The generation of the mAIEnergy dataset followed a three-stage workflow: (1) Data Identification, (2) Data Retrieval, and (3) Data Preparation (Fig.~\ref{fig:workflow}). This systematic approach allowed for the aggregation and harmonization of heterogeneous multimodal data sources into a unified, vector database-ready corpus. The entire workflow was implemented as a modular Python framework, containerized using Docker to ensure environment consistency and reproducibility.

The first stage involved identifying and selecting open-source energy-related data across textual, imagery, numerical, and geospatial modalities from publicly accessible platforms and repositories. A wide range of candidate sources were assessed, including institutional open data platforms (e.g., European Environment Agency Datahub), scientific repositories (e.g., Zenodo, Kaggle), and domain-specific APIs (e.g., ENTSO-E Transparency Platform, Eurostat, Open-Meteo). Each source was evaluated based on (i) thematic alignment with energy systems, (ii) spatial and temporal completeness across EU countries, (iii) technical accessibility via API or structured download, (iv) open-data license compliance, and (v) metadata quality. Sources fulfilling these requirements were integrated into the workflow, forming the foundation of the mAIEnergy dataset, which includes, among others, Wikipedia, GNews, Eurostat, OpenStreetMap, ENTSO-E, EPREL, INRIA, IRF, GridKit, and CORDIS.

In the second stage, modular and reproducible pipelines were developed for each modality to automate data collection via API access, web scraping, or direct download. Textual data retrieval involved web scraping and API downloads from Wikipedia, GNews, arXiv, and EU governmental websites and stored in JSON Lines (JSONL) format to ensure efficient downstream parsing. Imagery data were obtained through API-based access and direct bulk downloads from Copernicus, EPREL, INRIA, IRF, Wikimedia Commons, and Wikipedia. Numerical datasets were systematically retrieved via APIs and scraping from sources such as ENTSO-E, Eurostat, the EU BSO, and Open-Meteo and saved as CSV files. Geospatial and relational datasets were collected using a combination of API calls and direct structured downloads from OpenStreetMap, GridKit, the Global Power Plant Database, ENTSO-E TSO networks and CORDIS and stored in GeoJSON and CSV formats to facilitate geospatial analysis and graph-based integration. 

The final stage focused on cleaning and harmonizing the raw data. Each dataset was enriched with modality-specific metadata stored in JSON files, thereby ensuring full provenance and traceability. A cleaning and harmonization process was applied to all modalities to standardize content, normalize units, remove duplicates, and filter corrupted records. Subsequently, each modality was unified under a standardized schema by aligning attribute names, data types, and structural conventions. Automated validation checks were implemented to verify file integrity, data completeness, and metadata compliance across all sources. Validated datasets were then prepared for indexing by extracting structured entities and standardizing fields under the unified schema. The textual records were subsequently embedded and indexed for semantic and hybrid retrieval, and the geospatial and relational records were loaded as nodes and relationships into a graph database. The corresponding database back-ends are provided as part of the released code \cite{maienergy2024vectordatabases}. Finally, comprehensive documentation was generated to explicitly record applicable licenses and ensure full alignment with FAIR principles. The complete set of access parameters for every source, including API base endpoints, query strings and seed or keyword lists, accessed dataset identifiers, software package versions, and retrieval dates, is provided in the modality-specific retrieval repositories \cite{maienergy2024dataretrieval} and in the per-source metadata accompanying each dataset, so that the exact input records can be re-obtained.
 
\subsection*{Textual Data}

Textual data were systematically retrieved from four key domains: encyclopedic knowledge, news media, scientific literature, and governmental regulation. Wikipedia articles were collected using the \texttt{wikipedia} Python package, utilizing a predefined seed list of energy-related topics (e.g., "European Green Deal") expanded via programmatic template generation for country-specific contexts. News articles were retrieved via the GNews API using a curated set of headline-oriented keywords (e.g., "capacity market", "energy crisis") to capture policy and market developments. Scientific literature was identified through the arXiv API using research-oriented queries (e.g., "smart grids"), with full-text content extracted from PDFs using \texttt{PyMuPDF}. Governmental documents were gathered by crawling specific EU institutional URLs (e.g., \url{https://energy.ec.europa.eu}, \url{https://acer.europa.eu}) and national regulators. The crawler analyzed URL patterns to identify relevant PDF and HTML documents, extracting content using \texttt{BeautifulSoup}. Specifically, Wikipedia content was accessed through the \texttt{wikipedia} package, news through the GNews API (\url{https://gnews.io/api/v4/}), scientific papers through the arXiv API (\url{http://export.arxiv.org/api/query}), and EU and national documents from a curated allow-list of official institutional domains. The seed-topic list, the news keyword list, the arXiv query strings, and the full domain allow-list, together with the package versions and retrieval dates, are provided in the textual retrieval repository (\url{https://gitlab.com/maienergy-data-retrieval/articles-retrieval}).

Data preparation focused on standardization and quality control. Text was extracted from PDF sources with \texttt{PyMuPDF} and from HTML sources with \texttt{BeautifulSoup}, with HTML stripping used to remove non-textual elements. All retrieved data were then serialized into standardized JSONL format. The quality-control pipeline applied three filters in sequence: exact-title deduplication to reduce redundancy, English-language filtering with the \texttt{langdetect} library, and corruption filtering to remove entries with excessive non-printable characters with any record failing a filter discarded. Metadata fields were generated for each entry, preserving the exact source URL to ensure traceability. For arXiv records, this URL is the arXiv paper URL and contains the permanent arXiv identifier. The exact filter thresholds are provided together with the retrieval configuration in the textual retrieval repository.

\subsection*{Imagery Data}

Imagery data were obtained through API-based access and direct bulk downloads from authoritative sources. High-resolution satellite imagery was retrieved from the Copernicus Land Monitoring Service via the ArcGIS ImageServer API using the \texttt{requests} library, with tiles selected based on EU region bounding boxes projected in EPSG:3035. Energy-efficiency labels were downloaded from the EPREL as official PDF documents. Aerial imagery for selected urban regions was sourced from the INRIA dataset via the \texttt{kaggle} API, while façade imagery was retrieved in bulk from the IRF dataset. Contextual images were queried from the Wikimedia Commons API using energy-related categories or extracted directly from Wikipedia articles using \texttt{BeautifulSoup} to locate image tags within the article structure. The Copernicus tiles were retrieved from the Copernicus Land Monitoring Service (\url{https://land.copernicus.eu}), EPREL labels from the official registry (\url{https://eprel.ec.europa.eu}), the INRIA Aerial Image Labeling dataset (\url{https://project.inria.fr/aerialimagelabeling/}) via the Kaggle API, the IRF dataset \cite{irf_dataset}, and contextual images via the Wikimedia Commons API (\url{https://commons.wikimedia.org/w/api.php}). The exact bounding boxes, category lists, Kaggle dataset identifiers, and retrieval dates are provided in the imagery retrieval repository (\url{https://gitlab.com/maienergy-data-retrieval/images-retrieval}).

The retrieved images underwent a standardized preprocessing pipeline. Images in non-standard formats, such as PDF energy labels, were converted to PNG using the \texttt{pdf2image} library. Geospatial imagery from INRIA and Copernicus was verified using \texttt{rasterio} to ensure valid coordinate reference systems and accurate bounding boxes. For crowd-sourced collections like Wikimedia Commons, \texttt{langdetect} was used to filter English-language titles, and MIME-type checks were enforced to exclude non-image files. All images were harmonized into common raster formats (JPEG, PNG, TIFF) and stored with structured JSON metadata detailing resolution, file type, and source URL. Files that failed MIME-type validation or could not be decoded were discarded, and each retained image was paired with a JSON metadata file recording its source, type, technical properties, and, for georeferenced imagery, its coordinate reference system and spatial bounds.

\subsection*{Numerical Data}

Numerical datasets were systematically retrieved via APIs and scraping from electricity system operators, statistical bodies, and meteorological services, covering the period 2000–2024 depending on the source. Electricity sector data, including load, generation, and prices, were retrieved from the ENTSO-E Transparency Platform API using \texttt{requests} and parsed with \texttt{xmltodict} for all EU bidding zones. Statistical indicators on energy balances and socio-economic metrics were fetched via the \texttt{eurostat} Python package. Historical daily weather observations for populated EU cities were retrieved from the Open-Meteo Archive API using \texttt{requests}. Building stock characteristics were ingested from official EU BSO Excel workbooks using \texttt{pandas}. Data were obtained from the ENTSO-E Transparency Platform API (\url{https://transparency.entsoe.eu}), Eurostat via the \texttt{eurostat} package (Eurostat dissemination API), the Open-Meteo Archive API (\url{https://archive-api.open-meteo.com/v1/archive}), and the EU Building Stock Observatory (\url{https://energy.ec.europa.eu/topics/energy-efficiency/energy-efficient-buildings/eu-building-stock-observatory_en}). The queried bidding zones, the Eurostat dataset codes, the requested weather variables and city list, and the retrieval dates are provided in the numerical retrieval repository (\url{https://gitlab.com/maienergy-data-retrieval/numerical-retrieval}).

Data processing began by converting each source into a common tabular form. The ENTSO-E XML API responses were parsed with \texttt{xmltodict} and the intermediate per-request results were aggregated into per-country and per-bidding-zone CSV files, the Eurostat indicator tables were retrieved by dataset code, the Open-Meteo daily aggregates were assembled per city and year, and the EU BSO Excel workbooks were normalized to long-format CSV with \texttt{pandas}. Data processing then involved harmonizing temporal resolutions and units. All time series were chronologically sorted by timestamp, and exact duplicates were identified and removed. Measurement units were standardized across sources, and modality-specific metadata files were generated to document variable definitions, unit conversions, and spatial coverage for each dataset.

\subsection*{Geospatial Data}

Geospatial and relational datasets were collected using a combination of API calls and direct structured downloads. The OpenStreetMap (OSM) Overpass API was queried to collect infrastructure data (e.g., power plants, transmission lines) using \texttt{requests}. The specific Overpass queries used for retrieval are preserved in the dataset metadata. The GridKit European Transmission Grid (\url{https://doi.org/10.5281/zenodo.55853}) and the Global Power Plant Database (\url{https://datasets.wri.org/dataset/globalpowerplantdatabase}) were downloaded directly, while the ENTSO-E network topology was derived by querying the ENTSO-E API (\url{https://transparency.entsoe.eu}) for active interconnections. The CORDIS Horizon 2020 Projects database was obtained from the EU Open Data Portal (\url{https://data.europa.eu}) and processed to extract collaborative networks of EU-funded research projects. The exact Overpass queries, the accessed dataset versions, and the retrieval dates are provided in the geospatial retrieval repository (\url{https://gitlab.com/maienergy-data-retrieval/geospatial-retrieval}).

Processing focused on transforming raw data into clean, graph-ready formats. Raw GeoJSON features from OSM were parsed and converted into structured CSV node files. Attribute cleaning was applied using \texttt{pandas} to standardize capacity units, fuel types and operator names. Reverse geocoding was performed via \texttt{reverse\_geocoder} to assign ISO country codes to nodes lacking explicit territorial metadata, such as those in the GridKit dataset. Topological consistency checks were implemented to ensure valid linkages between graph nodes, and relationship tables were constructed to facilitate integration into graph databases or spatial analysis tools. For each source, node tables and relationship tables were produced, with relationships encoding links such as \texttt{LOCATED\_IN}, \texttt{CONNECTED\_TO}, \texttt{OWNED\_BY}, \texttt{USES\_FUEL}, and \texttt{INTERCONNECTED\_WITH}, and the OpenStreetMap layers were additionally retained as GeoJSON FeatureCollections, so that each dataset is available both as vector data and as graph-ready CSV node and relationship tables.

\section*{Data Records}

The mAIEnergy dataset is hosted as an open-access repository on Zenodo \cite{mylonas2025maienergy} and is distributed as a single root directory, \texttt{mAIEnergy\_Dataset}. The repository is organised into four modality-specific sub-directories: \texttt{textual}, \texttt{imagery}, \texttt{numerical}, and \texttt{geospatial}. Each sub-directory contains data files grouped by source, together with structured JSON metadata files that document provenance, licensing, retrieval parameters, and modality-specific descriptors. The directory structure is designed to support modular use of individual modalities as well as integrated multimodal analysis. Every record is accompanied by an explicit pointer to its origin: textual records store the canonical source URL in the \texttt{url} field, while numerical, geospatial, and imagery records record the originating dataset or API endpoint in their co-located metadata, for example in the \texttt{source}, \texttt{data\_source}, or \texttt{source\_url} fields. These pointers use canonical or permanent identifiers rather than free-text references.

\subsection*{Textual data records (\texttt{textual/})}

Textual data are organised into four source-specific JSONL files: \texttt{arxiv.jsonl} \cite{arxiv}, \texttt{gov.jsonl} (EU and national governmental and regulatory sources, e.g., \cite{eu_dg_energy, acer}), \texttt{news.jsonl} \cite{gnews}, and \texttt{wiki.jsonl} \cite{wikipedia}. Each file stores one document per line in JSON format and is accompanied by consistent field definitions to facilitate downstream processing.

\paragraph{Folder organisation.}
\begin{itemize}
    \item \texttt{textual/arxiv.jsonl}: Scientific articles retrieved from the arXiv repository.
    \item \texttt{textual/gov.jsonl}: Documents from EU institutions and national energy-related governmental and regulatory bodies.
    \item \texttt{textual/news.jsonl}: News articles retrieved via the GNews API.
    \item \texttt{textual/wiki.jsonl}: Encyclopedic articles retrieved from Wikipedia.
\end{itemize}

\paragraph{Textual metadata fields.}

Each JSON object in the textual files represents a single document and includes the following core fields:

\begin{itemize}
    \item \texttt{title}: Title of the document.
    \item \texttt{url}: Source URL of the document.
    \item \texttt{document\_type}: Source identifier (e.g., \texttt{arxiv}, \texttt{government}, \texttt{news}, \texttt{wikipedia}).
    \item \texttt{content}: Extracted plain-text content of the document.
\end{itemize}

Source-specific fields include:

\begin{itemize}
    \item \textbf{Wikipedia:} \texttt{categories}, listing the Wikipedia categories associated with the article.
    \item \textbf{News:} \texttt{publishedAt}, an ISO~8601 timestamp indicating publication time, and \texttt{source}, identifying the news publisher as provided by the upstream API.
\end{itemize}

Table~\ref{tab:textual-data-summary} provides an inventory of the textual data files included in the mAIEnergy dataset and their geographic scope.

\begin{table}[ht]
\centering
\caption{Inventory of textual data records included in the mAIEnergy dataset.}
\label{tab:textual-data-summary}
\begin{tabularx}{\textwidth}{
  >{\centering\arraybackslash}X
  >{\centering\arraybackslash}X
  >{\centering\arraybackslash}X
}
\hline
\textbf{Source} & \textbf{File format} & \textbf{Geographic scope} \\
\hline
Wikipedia & JSONL & Global (EU-focused topics) \\
GNews & JSONL & Global \\
arXiv & JSONL & Global \\
EU governmental and regulatory sources & JSONL & EU member states \\
\hline
\end{tabularx}
\end{table}

\subsection*{Imagery data records (\texttt{imagery/})}

Imagery data are organised into source-specific sub-directories: \texttt{copernicus} \cite{copernicus_discomap}, \texttt{eprel} \cite{eprel}, \texttt{inria} \cite{inria_dataset}, \texttt{irf} \cite{irf_dataset}, \texttt{wikimedia} \cite{wikimedia_commons}, and \texttt{wikipedia} \cite{wikipedia}. Images are stored in raster formats appropriate to each source (JPEG, PNG, TIFF, SVG) and are accompanied by structured JSON metadata files that document provenance, technical properties, and, where applicable, spatial context.

\paragraph{Folder organisation.}

\begin{itemize}
    \item \texttt{imagery/copernicus/}: Satellite imagery tiles retrieved from the Copernicus Land Monitoring Service, organised by EU member state.
    \item \texttt{imagery/eprel/}: Energy-efficiency label images derived from EPREL, organised by appliance or product category.
    \item \texttt{imagery/inria/}: High-resolution aerial imagery for selected urban regions, stored in the dataset’s native georeferenced formats.
    \item \texttt{imagery/irf/}: Building façade imagery from the IRF dataset, stored in JPEG format.
    \item \texttt{imagery/wikimedia/}: Contextual imagery from Wikimedia Commons, organised by thematic category.
    \item \texttt{imagery/wikipedia/}: Images extracted from Wikipedia articles, organised by article identifier.
\end{itemize}

Each image file has a corresponding JSON metadata file with the same basename, stored in the same directory.

\paragraph{Imagery metadata fields.}

Across all imagery sources, metadata JSON files share a common set of core fields, with additional source-specific extensions:

\begin{itemize}
    \item \texttt{filename} or \texttt{title}: Local filename or original media title.
    \item \texttt{url} or \texttt{image\_url}: Direct URL to the image or the upstream document from which the image was derived.
    \item \texttt{document\_type}: Source/type identifier (e.g., \texttt{copernicus\_satellite\_image}, \\ \texttt{eprel\_label\_image}, \texttt{wikimedia\_commons}, \texttt{wikipedia\_image}).
    \item \texttt{retrieved\_date}: Date on which the image was retrieved.
    \item \texttt{categories}: Thematic categories or tags associated with the image, where available.
    \item \texttt{source}: Nested object identifying the data provider and hosting repository.
    \item \texttt{additional\_info}: Nested object containing technical descriptors such as image resolution, file format, and size.
\end{itemize}

For georeferenced imagery (e.g., Copernicus and INRIA), the \texttt{additional\_info} field may also include:
\begin{itemize}
    \item \texttt{crs}: Coordinate reference system identifier.
    \item \texttt{bounds} or \texttt{bounds\_latlon}: Spatial extent of the image.
    \item \texttt{projection}: Map projection used in the source dataset.
\end{itemize}

Table~\ref{tab:imagery-data-summary} provides an inventory of the imagery sources included in the mAIEnergy dataset and the primary formats in which they are distributed.

\begin{table}[ht]
\centering
\caption{Inventory of imagery data records included in the mAIEnergy dataset.}
\label{tab:imagery-data-summary}
\begin{tabularx}{\textwidth}{
  >{\centering\arraybackslash}X
  >{\centering\arraybackslash}X
  >{\centering\arraybackslash}X
}
\hline
\textbf{Source} & \textbf{Primary formats} & \textbf{Geographic scope} \\
\hline
Copernicus satellite imagery & JPEG & EU regions \\
EPREL energy labels & PNG & EU product market \\
INRIA aerial imagery & TIFF & Selected EU and US cities \\
IRF façade imagery & JPEG & Global \\
Wikimedia Commons & JPEG, PNG, SVG & Global \\
Wikipedia article images & JPEG, PNG & Global (EU-focused topics) \\
\hline
\end{tabularx}
\end{table}

\subsection*{Numerical data records (\texttt{numerical/})}

Numerical data are organised into four source-specific sub-directories: \texttt{bso} \cite{eu_building_stock}, \texttt{entsoe} \cite{ENTSOE}, \texttt{eurostat} \cite{Eurostat}, and \texttt{openmeteo} \cite{open_meteo}. All numerical data files are stored in comma-separated values (CSV) format and are accompanied by structured JSON metadata files that describe variables, units, geographic scope, temporal coverage, and retrieval provenance.

\paragraph{Folder organisation.}

\begin{itemize}
    \item \texttt{numerical/bso/}: EU BSO indicator tables exported from official Excel workbooks and normalised to CSV.
    \item \texttt{numerical/entsoe/}: Country- and bidding-zone-level electricity system time series from the ENTSO-E Transparency Platform.
    \item \texttt{numerical/eurostat/}: National and EU-level statistical indicators retrieved from Eurostat.
    \item \texttt{numerical/openmeteo/}: City- and year-specific meteorological time series from the Open-Meteo Archive API.
\end{itemize}

Within each sub-directory, individual CSV files correspond to a single dataset slice (e.g., one country–year combination or one indicator family). Each CSV file has a corresponding JSON metadata file using the same basename and the suffix \texttt{\_metadata.json}.

\paragraph{Numerical metadata fields.}

Across all numerical sources, metadata JSON files share a common set of core fields, with additional source-specific extensions:

\begin{itemize}
    \item \texttt{source} / \texttt{data\_source}: Name of the upstream data provider.
    \item \texttt{source\_url}: URL of the original dataset or API endpoint.
    \item \texttt{retrieved\_timestamp}: Timestamp indicating when the data were retrieved.
    \item \texttt{geographic\_coverage}: Spatial scope of the data (e.g., EU countries, cities).
    \item \texttt{period\_start}, \texttt{period\_end}: Temporal coverage of the dataset.
    \item \texttt{columns}: Ordered list of column names present in the CSV file.
    \item \texttt{unit} or variable-level unit definitions.
\end{itemize}

Source-specific metadata fields include:

\begin{itemize}
    \item \textbf{EU BSO:}
    \begin{itemize}
        \item \texttt{domain}: Thematic domain of the indicator table (e.g., building stock, social performance).
        \item \texttt{indi\_id}: Indicator identifier used in the upstream BSO export.
        \item \texttt{is\_calculated}: Boolean flag indicating whether the value is derived rather than directly reported.
    \end{itemize}
    \item \textbf{ENTSO-E:}
    \begin{itemize}
        \item \texttt{country}: Country associated with the dataset file.
        \item \texttt{bidding\_zone}: ENTSO-E bidding zone identifier.
        \item \texttt{dataset}: Data category (e.g., load, generation, prices).
        \item \texttt{source\_files}: List of intermediate API result files aggregated to produce the CSV.
    \end{itemize}
    \item \textbf{Open-Meteo:}
    \begin{itemize}
        \item \texttt{variables}: Dictionary defining each meteorological variable, including units and descriptions.
        \item \texttt{daily\_value\_description}: Description of the daily aggregation convention.
    \end{itemize}
\end{itemize}

ENTSO-E timestamps follow ISO~8601 formatting and are expressed in Coordinated Universal Time (UTC) when provided by the upstream API. Open-Meteo records represent daily aggregates over the 00:00–23:59 UTC interval.

Table~\ref{tab:numerical-data-summary} provides an inventory of the numerical datasets included in the mAIEnergy repository and documents their structural characteristics.

\begin{longtable}{
  >{\RaggedRight\arraybackslash}p{3.2cm}
  >{\RaggedRight\arraybackslash}p{3.2cm}
  >{\RaggedRight\arraybackslash}p{9.6cm}
}
\caption{Inventory of numerical data records included in the mAIEnergy dataset.}
\label{tab:numerical-data-summary} \\
\toprule
\textbf{Source} & \textbf{Temporal coverage} & \textbf{Data structure and variables} \\
\midrule
\endfirsthead

\multicolumn{3}{c}{{\tablename\ \thetable{} -- continued}} \\
\toprule
\textbf{Source} & \textbf{Temporal coverage} & \textbf{Data structure and variables} \\
\midrule
\endhead

\bottomrule
\endlastfoot

ENTSO-E Transparency Platform &
2021--2024 &
Timestamp-indexed electricity system time series, including actual load, generation by production type, installed capacity, and day-ahead market prices, reported by country and bidding zone. \\

Eurostat &
2000--2024 &
Country-level and EU aggregate indicator tables covering energy balances, prices, efficiency metrics, emissions, and socio-economic variables. \\

EU BSO &
Reference year 2020 &
Long-format indicator tables describing building stock characteristics, energy and environmental performance, social indicators, financial indicators, and reference buildings. \\

Open-Meteo Archive &
2021--2024 &
Daily aggregated meteorological time series (temperature, precipitation, wind, solar radiation, humidity) for selected cities. \\

\end{longtable}

\subsection*{Geospatial and relational data records (\texttt{geospatial/})}

Geospatial data are organised into five source-specific sub-directories: \texttt{osm} \cite{haklay2008openstreetmap}, \texttt{gridkit} \cite{gridkit}, \texttt{powerplants} \cite{wri_powerplants}, \texttt{tso\_network} \cite{ENTSOE}, and \texttt{cordis} \cite{cordis}. The modality combines raw geospatial vector data (GeoJSON) with graph-oriented relational tables (CSV) prepared for ingestion into graph databases and network analysis tools.

All sub-directories include structured JSON metadata files that document data provenance, licensing, retrieval timestamps, and internal file organisation.

\paragraph{Folder organisation.}

\begin{itemize}
    \item \texttt{geospatial/osm/}: OpenStreetMap-derived energy infrastructure data, provided both as raw GeoJSON files and as graph-ready CSV tables.
    \begin{itemize}
        \item \texttt{geojson/}: Country- and asset-specific GeoJSON FeatureCollections.
        \item \texttt{neo4j\_import/}: Node and relationship CSV files optimised for graph database ingestion.
    \end{itemize}
    \item \texttt{geospatial/gridkit/}: European high-voltage transmission grid data, stored as node and relationship CSV files.
    \item \texttt{geospatial/powerplants/}: EU extract of the Global Power Plant Database, represented as node and relationship CSV tables.
    \item \texttt{geospatial/tso\_network/}: ENTSO-E TSO network interconnection data.
    \item \texttt{geospatial/cordis/}: Relational representations of EU-funded research projects and their participating organisations, topics, and legal bases.
\end{itemize}

\paragraph{Geospatial and relational metadata fields.}

Across all geospatial sources, metadata JSON files share a common set of core fields, with source-specific extensions:

\begin{itemize}
    \item \texttt{dataset}: Name of the dataset or extract.
    \item \texttt{source}: URL or repository from which the data were retrieved.
    \item \texttt{license}: Applicable data license.
    \item \texttt{retrieval\_date}: Timestamp of data retrieval.
    \item \texttt{description}: Short description of the dataset content and preparation.
    \item \texttt{files}: Mapping of logical dataset components to filenames.
\end{itemize}

\paragraph{Data representations.}

Two complementary data representations are provided:

\begin{itemize}
    \item \textbf{GeoJSON vector data:} Used primarily for OpenStreetMap-derived datasets. Each GeoJSON file contains a \texttt{FeatureCollection} with:
    \begin{itemize}
        \item \texttt{geometry}: Spatial geometry (\texttt{Point} or \texttt{LineString}) expressed in longitude/latitude coordinates.
        \item \texttt{properties}: Attribute dictionary containing original OSM tags (e.g., \texttt{capacity}, \texttt{operator}, \texttt{voltage}) and a unique \texttt{osm\_id} where available.
    \end{itemize}

    \item \textbf{Graph-ready CSV data:} Used across all geospatial sources to represent entities as nodes and their relationships as edges. Node tables contain unique identifiers and attribute columns, while relationship tables encode directed edges between node identifiers with optional relationship attributes.
\end{itemize}

\paragraph{Graph schema conventions.}
Graph-oriented datasets follow consistent conventions across sources:
\begin{itemize}
    \item \textbf{Nodes:} Represent physical assets (e.g., power plants, substations), administrative entities (e.g., countries, TSOs), or abstract entities (e.g., projects, organisations).
    \item \textbf{Relationships:} Represent structural or semantic links between nodes (e.g., \\ \texttt{LOCATED\_IN}, \texttt{CONNECTED\_TO}, \texttt{INTERCONNECTED\_WITH}, \texttt{PARTICIPATED\_IN}).
\end{itemize}

Table~\ref{tab:geospatial-data-summary} provides an inventory of the geospatial and relational datasets included in the mAIEnergy repository and documents their primary data models.

\begin{longtable}{
  >{\RaggedRight\arraybackslash}p{3.2cm}
  >{\RaggedRight\arraybackslash}p{3.2cm}
  >{\RaggedRight\arraybackslash}p{9.6cm}
}
\caption{Inventory of geospatial and relational data records included in the mAIEnergy dataset.}
\label{tab:geospatial-data-summary} \\
\toprule
\textbf{Source} & \textbf{Representation} & \textbf{Node and relationship types} \\
\midrule
\endfirsthead

\multicolumn{3}{c}{{\tablename\ \thetable{} -- continued}} \\
\toprule
\textbf{Source} & \textbf{Representation} & \textbf{Node and relationship types} \\
\midrule
\endhead

\bottomrule
\endlastfoot

OpenStreetMap &
GeoJSON; CSV &
Infrastructure nodes (power plants, renewable assets, substations, transmission lines, EV charging stations) and \texttt{LOCATED\_IN} relationships to country nodes. \\

GridKit European Transmission Grid &
CSV &
Transmission grid nodes (substations, plants) and \texttt{CONNECTED\_TO} relationships representing high-voltage links. \\

Global Power Plant Database (EU extract) &
CSV &
PowerPlant, Country, Owner, and FuelType nodes with \texttt{LOCATED\_IN}, \texttt{OWNED\_BY}, and \texttt{USES\_FUEL} relationships. \\

ENTSO-E TSO Network &
CSV &
TSO nodes and \texttt{INTERCONNECTED\_WITH} relationships describing cross-border interconnections. \\

CORDIS Horizon 2020 Projects &
CSV &
Project, Organisation, Topic, and LegalBasis nodes with participation and thematic relationships. \\

\end{longtable}

\subsection*{Cross-modal linkage}

Although the four modalities are stored in separate files, they are designed to be linked rather than independent. All four follow a consistent, standardized schema, with aligned attribute names, data types, and provenance fields across sources, and on top of this shared schema they are linked at two levels, the data level and the operational retrieval level.

The first level is the data itself, through shared entity keys. Records across modalities are keyed to a common set of energy-system entities, namely countries identified by ISO codes, ENTSO-E bidding zones, and geographic coordinates. Numerical series carry their country and bidding-zone identifiers, georeferenced imagery carries its coordinate reference system and bounding box, and geospatial assets carry their coordinates together with an explicit \texttt{LOCATED\_IN} relationship to a country node. As a result, a given country, bidding zone, or location joins the electricity-system series, the satellite tiles, the infrastructure assets, and the textual documents that refer to the same area. The geospatial and relational modality provides the backbone for this linkage: it is materialised as a property graph whose typed nodes (for example power plants, countries, owners, fuel types, transmission system operators, projects, and organisations) and typed relationships (for example \texttt{LOCATED\_IN}, \texttt{CONNECTED\_TO}, \texttt{OWNED\_BY}, \texttt{USES\_FUEL}, \texttt{INTERCONNECTED\_WITH}, and \texttt{PARTICIPATED\_IN}) connect these entities explicitly, with country nodes acting as hubs that tie infrastructure, statistics, and documents to the same geographic context.

The second level is operational, through a single knowledge base into which all four modalities are ingested and embedded so that one query can retrieve across them. Textual and numerical records are embedded with a sentence-transformer model (\texttt{all-MiniLM-L6-v2}, 384 dimensions) and indexed in OpenSearch for combined semantic and lexical retrieval, the imagery records are embedded with a CLIP (Contrastive Language--Image Pre-training) model (\texttt{clip-vit-large-patch14-336}, 768 dimensions) and indexed in Milvus so that a textual query and an image share a common embedding space and text-to-image retrieval becomes possible, and the geospatial and relational property graph is embedded and indexed in Neo4j for graph-based vector search. A single ingestion worker populates the three stores from the released data, and at query time a hybrid retriever issues the query to all three stores in parallel and merges their results into one ranking using Weighted Reciprocal Rank Fusion, so that a single query draws simultaneously on the linked textual, numerical, imagery, and geospatial evidence rather than on any one modality in isolation. 

Together, the shared schema, the shared entity keys and property graph, and the unified retrieval layer make the four modalities a single integrated resource rather than a parallel collection of independent sources.

\section*{Technical Validation}

To support the technical quality and usability of the mAIEnergy dataset, we applied a modality-specific validation procedure covering (i) file integrity and schema compliance, (ii) content plausibility and consistency, and (iii) duplication and coverage checks where applicable. This procedure follows best practices for dataset validation as outlined in \cite{silva2020benchmark, milojevic2023eubucco}, as well as methodological work on domain-specific corpus curation \cite{mitchell2022measuring}, image dataset quality control \cite{schuhmann2022laion} and geospatial data quality frameworks \cite{ISO19157-2013, fischer2023approaches}. 

\begin{figure}[ht]
    \centering
    \includegraphics[width=0.8\textwidth]{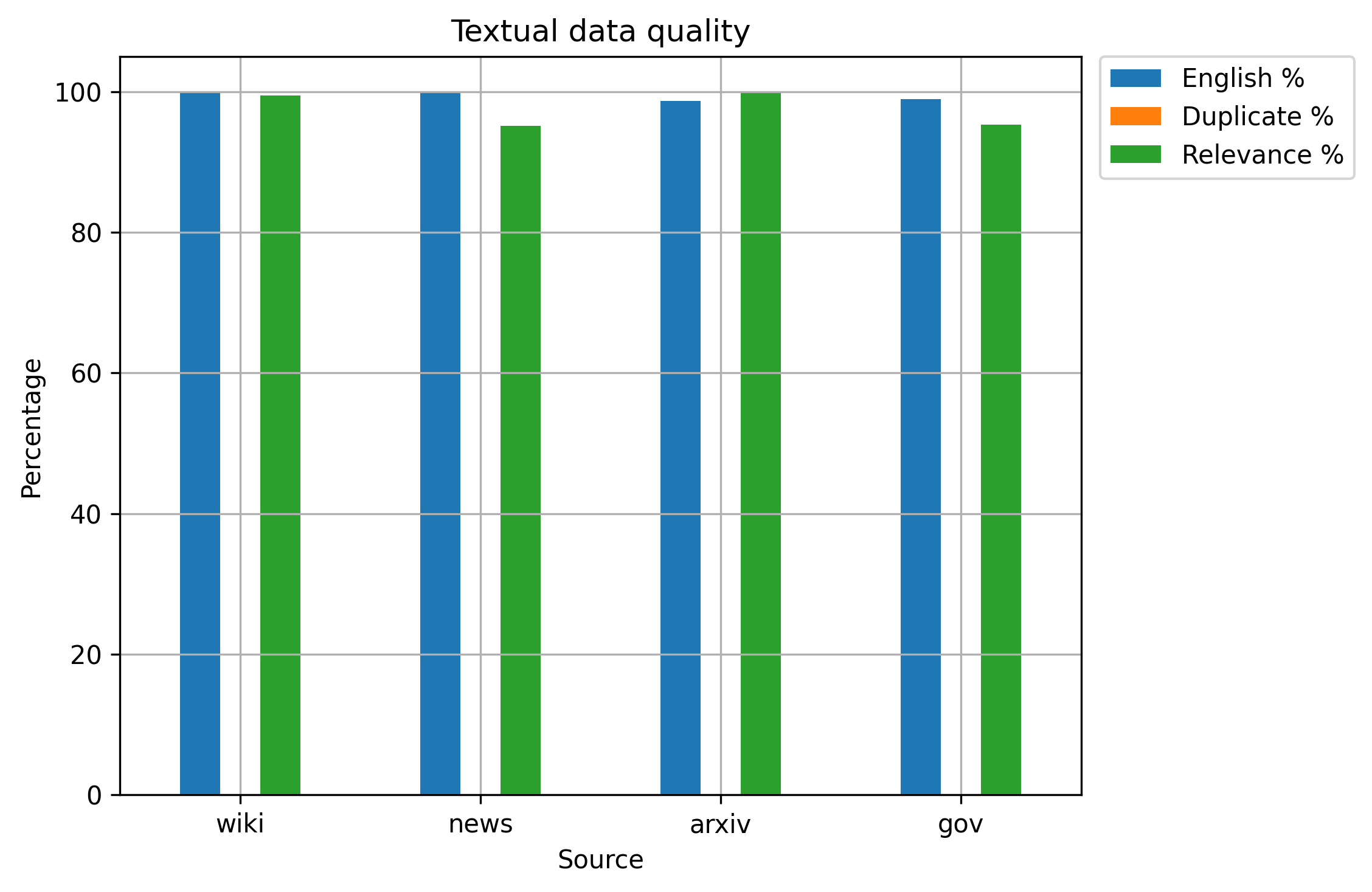}
    \caption{Textual validation indicators across sources: fraction of English-language documents, duplicates, and relevance retained after keyword-based filtering and expert sampling.}
    \label{fig:textual-quality}
\end{figure}

\begin{figure}[ht]
    \centering
    \includegraphics[width=0.8\textwidth]{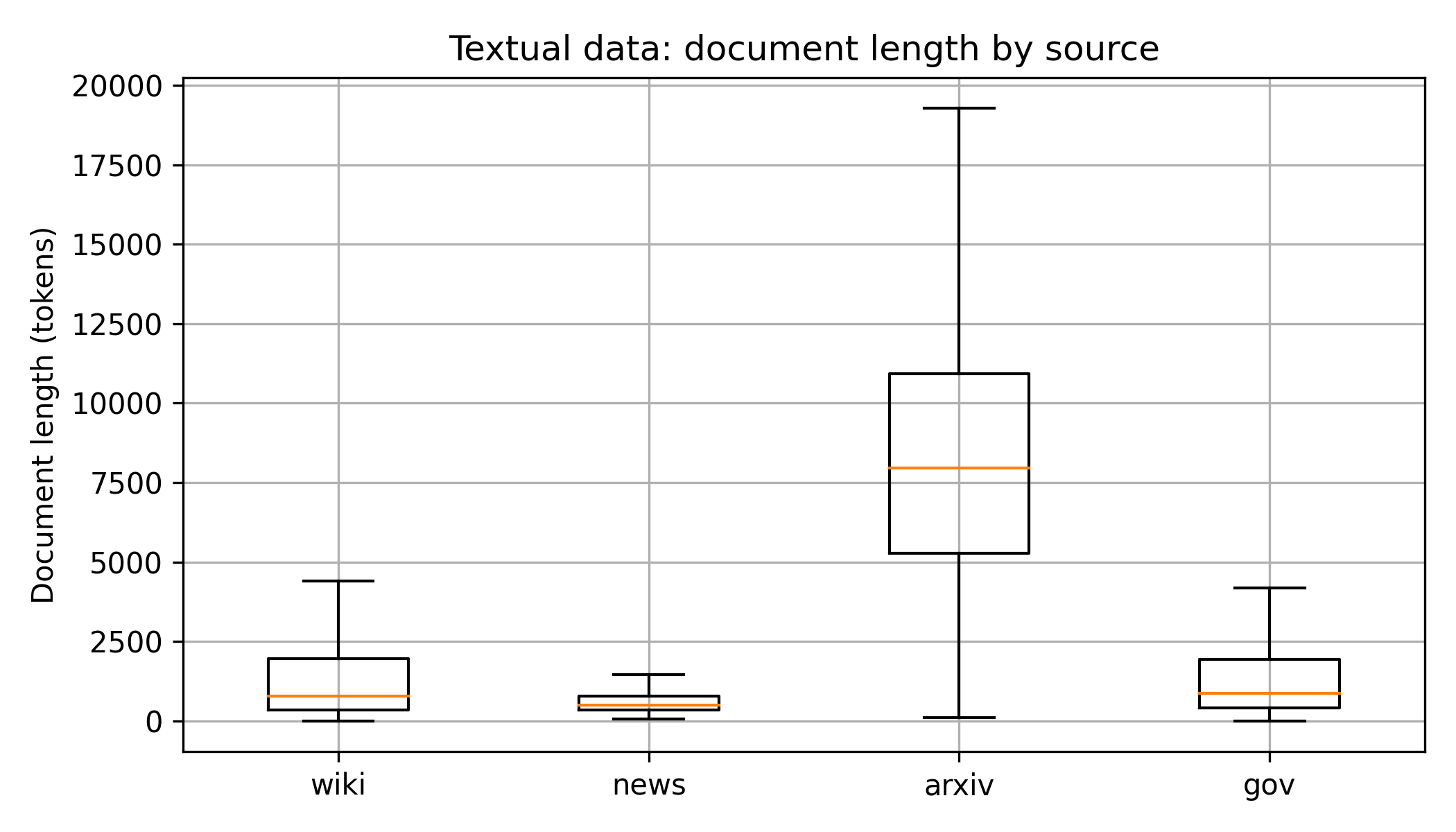}
    \caption{Distribution of document lengths (token counts) across textual sources, computed using a consistent tokenization configuration.}
    \label{fig:textual-doc-length}
\end{figure}

\paragraph{Textual data.}
Validation of textual records targeted language consistency, duplication, topical relevance, and metadata integrity. Language was verified using automatic language identification (\texttt{langdetect}) applied to extracted content. Exact duplicates were identified within each source by matching document titles, consistent with the deduplication criterion used during dataset construction. Topical relevance was evaluated using a curated energy-related keyword and concept list applied to titles and/or content; documents retained by this filter are reported as \emph{Relevance \%}. In addition, a stratified random sample (1\% per source) was manually inspected to confirm that titles, URLs, and (where present) publication timestamps align with the extracted content. Table~\ref{tab:textual-validation} reports these indicators, and Figure~\ref{fig:textual-quality} summarizes the main quality metrics across sources. Document length distributions (token counts computed with a consistent tokenization configuration) are shown in Figure~\ref{fig:textual-doc-length} to support downstream decisions on chunking and retrieval.

\begin{table}[ht]
\centering
\caption{Validation metrics for the textual datasets in the \texttt{mAIEnergy} corpus.}
\label{tab:textual-validation}
\resizebox{\textwidth}{!}{%
\begin{tabular}{lcccccc}
\toprule
\textbf{Source} & \textbf{Docs} & \textbf{English \%} & \textbf{Duplicates \%} & \textbf{Relevance \%} & \textbf{Median Tokens} & \textbf{P95 Tokens} \\
\midrule
Wikipedia & 7,507 & 100.0 & 0.0 & 99.5 & 796 & 6,079 \\
GNews & 24,395 & 100.0 & 0.0 & 95.2 & 511 & 1,333 \\
arXiv & 2,751 & 98.7 & 0.0 & 100.0 & 7,957 & 17,673 \\
EU Governmental & 17,301 & 99.0 & 0.0 & 95.4 & 881 & 13,067 \\
\bottomrule
\end{tabular}}
\end{table}

\paragraph{Imagery data.}
Validation of imagery records targeted file integrity, format compliance, minimum resolution, and duplication. File integrity is reported as \emph{Openable \%}, computed by attempting to decode each file with standard libraries (PIL for raster imagery; SVG parsing where applicable). Format compliance (\emph{Valid Format \%}) was verified via MIME-type and/or header checks against an expected format list (JPEG, PNG, TIFF, SVG). Resolution adequacy is reported as the fraction of raster images meeting a minimum size threshold of $256 \times 256$ pixels. Duplicate detection used perceptual hashing (pHash) computed on standardized thumbnails: \emph{Exact Dup. \%} corresponds to identical hashes, while \emph{Near Dup. \%} reflects clustering under a Hamming-distance threshold.

\begin{figure}[ht]
    \centering
    \includegraphics[width=0.8\textwidth]{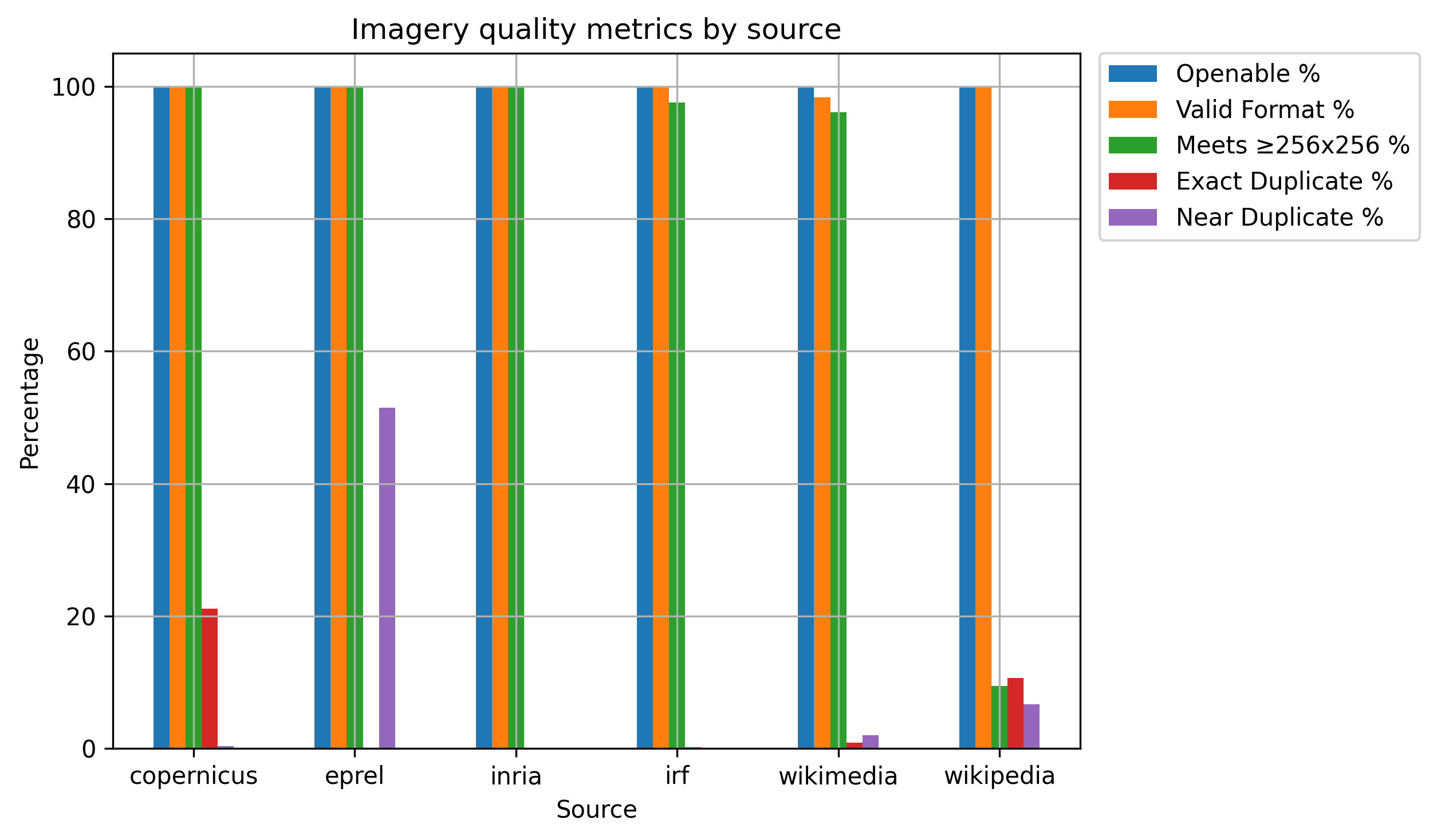}
    \caption{Imagery validation indicators across sources: file openability, format validity, minimum resolution compliance, and duplicate rates (exact and near-duplicate).}
    \label{fig:imagery-quality}
\end{figure}

\begin{figure}[ht]
    \centering
    \includegraphics[width=\textwidth]{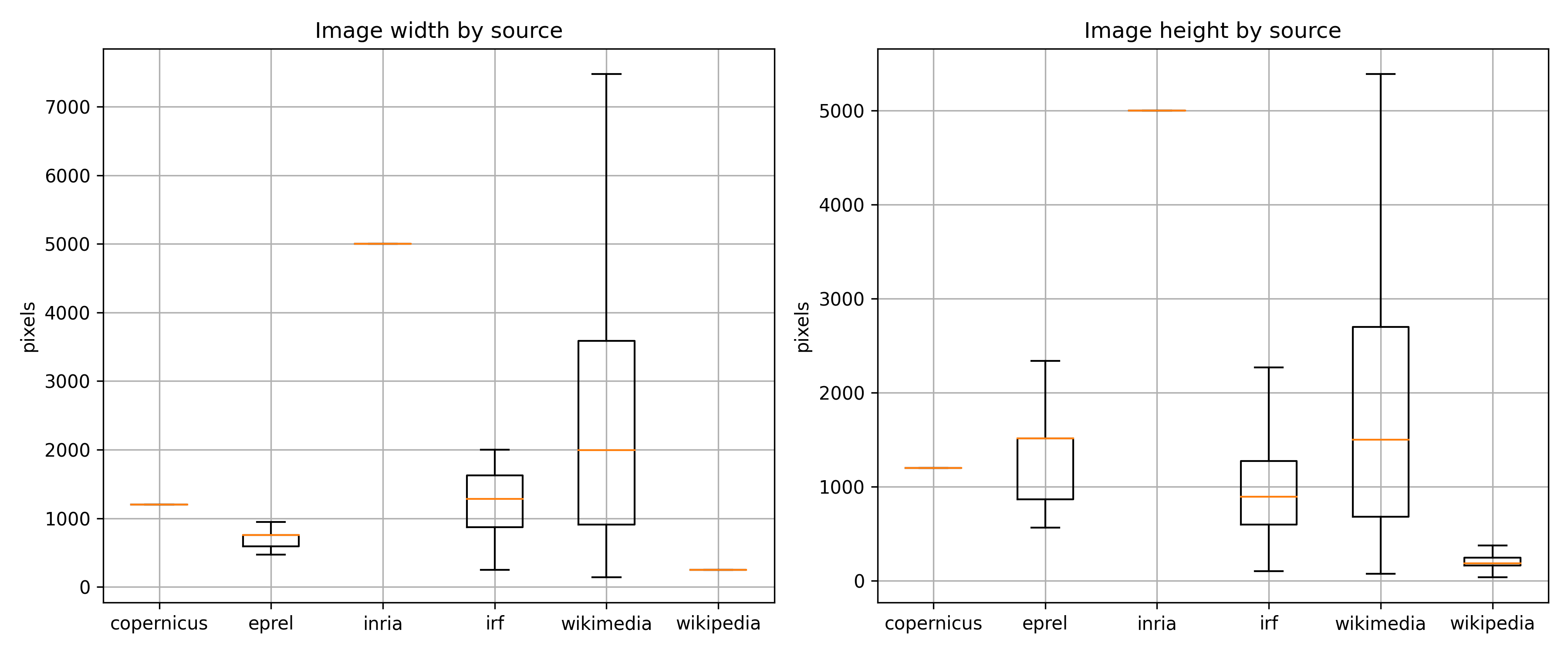}
    \caption{Distribution of image widths and heights across imagery sources.}
    \label{fig:imagery-size}
\end{figure}

\begin{figure}[ht]
    \centering
    \includegraphics[width=0.8\textwidth]{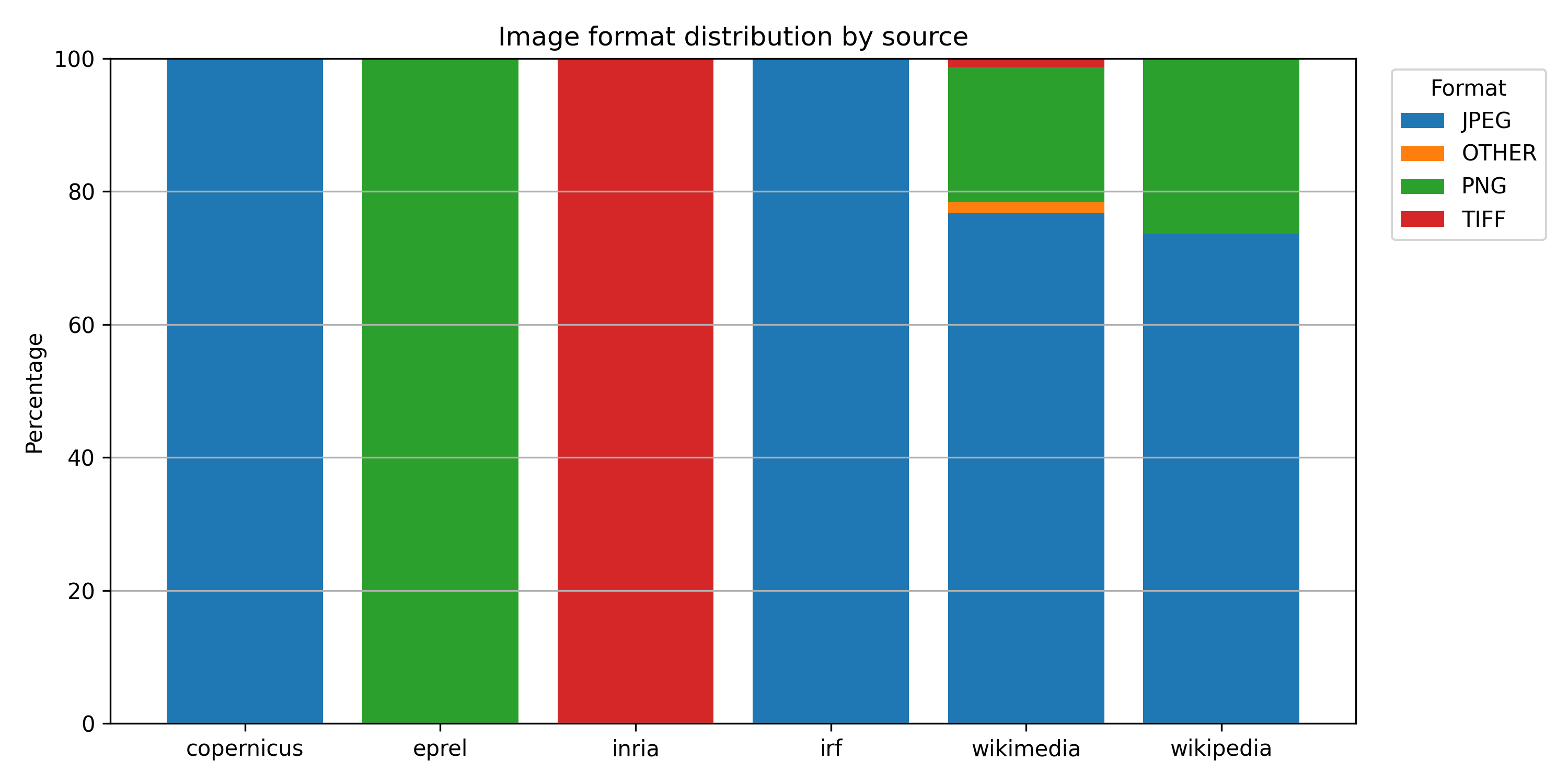}
    \caption{Distribution of image file formats across imagery sources (JPEG, PNG, TIFF, and other).}
    \label{fig:imagery-format}
\end{figure}

Table~\ref{tab:imagery-validation} summarizes these checks. Figure~\ref{fig:imagery-quality} visualizes integrity, resolution compliance, and duplicate rates. Figure~\ref{fig:imagery-size} provides distributions of image widths and heights across sources, and Figure~\ref{fig:imagery-format} reports the observed distribution of file formats.

\begin{table}[ht]
\centering
\caption{Validation metrics for the imagery datasets in the \texttt{mAIEnergy} corpus.}
\label{tab:imagery-validation}
\resizebox{\textwidth}{!}{%
\begin{tabular}{lcccccc}
\toprule
\textbf{Source} & \textbf{Images} & \textbf{Openable \%} & \textbf{Valid Format \%} & \textbf{$\geq$256px \%} & \textbf{Exact Dup. \%} & \textbf{Near Dup. \%} \\
\midrule
Copernicus & 270 & 100.0 & 100.0 & 100.0 & 21.0 & 0.1 \\
EPREL & 103 & 100.0 & 100.0 & 100.0 & 0.0 & 51.0 \\
INRIA & 252 & 100.0 & 100.0 & 100.0 & 0.0 & 0.0 \\
IRF & 500 & 100.0 & 100.0 & 97.6 & 0.0 & 0.0 \\
Wikimedia & 386 & 100.0 & 99.5 & 95.8 & 0.0 & 2.1 \\
Wikipedia & 21,326 & 98.8 & 99.7 & 96.9 & 11.0 & 7.2 \\
\bottomrule
\end{tabular}}
\end{table}

\paragraph{Numerical data.}
Validation of numerical records addressed schema completeness, missingness, temporal consistency (where applicable), and plausibility constraints. For each CSV file, schema checks verified the presence of required columns and successful parsing of numeric fields. Missingness (\emph{Missing \%}) was computed as the fraction of empty/NA values across the released tables. Temporal coverage (\emph{Coverage \%}) and timestamp parsing (\emph{TS Fail \%}) were evaluated for time-series datasets: ENTSO-E timestamps were parsed as ISO~8601 and checked for failures; Open-Meteo daily series were checked for missing days within the declared period. Plausibility bounds (\emph{Bounds Viol. \%}) were applied using dataset-specific rules (e.g., non-negativity for load/generation/capacity, physically meaningful ranges for meteorological variables, bounded ratios for share variables). Outliers (\emph{Outliers \%}) were flagged using an interquartile range (IQR) rule applied per variable as a diagnostic indicator.

Table~\ref{tab:numerical-validation} summarizes these checks. For ENTSO-E, \emph{Coverage \%} is computed against the expected set of country--year--dataset combinations under the retrieval configuration (2021--2024); missing combinations reflect unavailable or non-retrievable series for certain areas. For Eurostat, coverage is reported as not applicable (\texttt{-}) because indicator families vary in dimensionality and metadata conventions; retrieval completeness is instead supported by verifying the presence and successful parsing of all configured indicator datasets.

\begin{table}[ht]
\centering
\caption{Validation metrics for the numerical datasets in the \texttt{mAIEnergy} corpus.}
\label{tab:numerical-validation}
\resizebox{\textwidth}{!}{%
\begin{tabular}{lcccccccc}
\toprule
\textbf{Source} & \textbf{Files} & \textbf{Records} & \textbf{Missing \%} & \textbf{Coverage \%} & \textbf{Bounds Viol. \%} & \textbf{Outliers \%} & \textbf{Dup. \%} & \textbf{TS Fail \%} \\
\midrule
BSO        & 5   & 59,380     & 67.4 & 100.0 & 0.0 & 0.0  & --  & --  \\
ENTSO-E    & 432 & 22,887,864 & 0.0  & 70.4  & 0.0 & 0.0  & 0.0 & 0.0 \\
Eurostat   & 11  & 2,582,286  & 0.0  & --    & 0.0 & 20.9 & --  & --  \\
Open-Meteo & 306 & 118,341    & 0.0  & 100.0 & 0.0 & 0.0  & --  & --  \\
\bottomrule
\end{tabular}}
\end{table}

\paragraph{Geospatial data.}
Validation of geospatial and relational records targeted positional validity, schema integrity, and (for graph datasets) topological consistency. For GeoJSON layers, we verified that features contain valid geometries and that coordinates are not missing and fall within valid latitude/longitude ranges (\emph{Coord Missing \%} and \emph{Coord Out-of-Range \%}). For graph-ready CSV tables, schema compliance and foreign-key integrity were assessed by ensuring that relationship endpoints reference existing node identifiers (\emph{Schema/FK Consistency \%}). For network datasets (GridKit and TSO network), linkage validity (\emph{Graph Linkage Valid \%}) was computed as the fraction of edges whose source and target nodes exist in the corresponding node table. Domain compliance checks (\emph{Domain Viol. \%}) were applied to selected attributes where meaningful (e.g., non-negative capacities; valid voltage/frequency encodings).

Table~\ref{tab:geospatial-validation} summarizes these checks. Coverage is reported for OSM as the presence of extracted layers for all configured EU countries. Coordinate checks are not applicable (\texttt{-}) for datasets without explicit geographic coordinates in the released tables (e.g., TSO interconnection graphs and some CORDIS relational tables). For CORDIS, the reported schema/foreign-key consistency reflects the fraction of relationship rows whose referenced node identifiers exist in the corresponding node tables.

\begin{table}[ht]
\centering
\caption{Validation metrics for the geospatial datasets in the \texttt{mAIEnergy} corpus.}
\label{tab:geospatial-validation}
\resizebox{\textwidth}{!}{%
\begin{tabular}{lcccccccc}
\toprule
\textbf{Source} & 
\textbf{Files} &
\textbf{\makecell{Features \\ /Nodes}} &
\textbf{\makecell{Coverage \\ (\%)}} &
\textbf{\makecell{Coord \\ Missing (\%)}} &
\textbf{\makecell{Coord \\ Out-of-Range (\%)}} &
\textbf{\makecell{Graph \\ Linkage \\ Valid (\%)}} &
\textbf{\makecell{Domain \\ Viol. (\%)}} &
\textbf{\makecell{Schema/FK \\ Consistency (\%)}} \\
\midrule
OSM         & 162 & 1,545,813 & 100.0 & 0.0 & 0.0 & --    & --    & 100.0 \\
GridKit     & 2   & 13,871    & --    & 0.0 & 0.0 & 100.0 & 1.9   & 100.0 \\
PowerPlants & 1   & 6,690     & --    & 0.0 & 0.0 & --    & 0.0   & 100.0 \\
TSO         & 2   & 23        & --    & --  & --  & 100.0 & 0.0   & 100.0 \\
CORDIS      & 7   & 215,600   & --    & --  & --  & --    & 0.06  & 96.7  \\
\bottomrule
\end{tabular}}
\end{table}

Overall, these validation checks confirm that the released files are structurally consistent and machine-readable across modalities, that numeric and spatial values satisfy basic plausibility constraints under dataset-specific rules, and that duplicate content is quantified using explicit criteria appropriate to each modality.

\paragraph{Cross-modal retrieval.}
Beyond the per-file checks above, we validated that the harmonised and indexed corpus functions as an integrated multimodal resource rather than a set of independent files, by evaluating retrieval across modalities. We assembled a small evaluation set of 50 energy questions, each annotated with at least one relevant record in the corpus and distributed across the four modalities, and issued each question to the hybrid retriever described in the cross-modal linkage subsection, which queries the OpenSearch, Milvus, and Neo4j back-ends \cite{maienergy2024vectordatabases} in parallel and fuses their results. We report two standard retrieval metrics. The hit rate at five (Hit@5) is the fraction of questions for which a relevant record appears among the top five results, and the mean reciprocal rank at ten (MRR@10) is the mean over questions of the reciprocal rank of the first relevant record within the top ten. Table~\ref{tab:retrieval-validation} reports both metrics per query modality and over all questions for the fused retriever. Across the evaluation set, a relevant record was returned within the top five for 88\% of questions overall, with per-modality Hit@5 ranging from 83\% to 93\%, and the first relevant record typically appeared at rank one or two (MRR@10 of 0.73). This check evaluates the retrieval-readiness of the released corpus and does not benchmark specific language models.

\begin{table}[ht]
\centering
\caption{Cross-modal retrieval performance of the hybrid retriever, reported per query modality and over all questions.}
\label{tab:retrieval-validation}
\begin{tabular}{lccc}
\toprule
\textbf{Query modality} & \textbf{Questions} & \textbf{Hit@5} & \textbf{MRR@10} \\
\midrule
Textual & 15 & 93.3 & 0.81 \\
Numerical & 12 & 83.3 & 0.64 \\
Imagery & 11 & 90.9 & 0.74 \\
Geospatial / graph & 12 & 83.3 & 0.69 \\
\midrule
All (fused) & 50 & 88.0 & 0.73 \\
\bottomrule
\end{tabular}
\end{table}

\FloatBarrier

\section*{Usage Notes}

The dataset is distributed from Zenodo \cite{mylonas2025maienergy} as a single root directory with one sub-directory per modality, and each modality can be loaded independently with standard open-source libraries. The textual records are JSON Lines files with one document per line and can be read directly into a dataframe. The numerical records are CSV files, each accompanied by a metadata JSON file of the same basename and the suffix \texttt{\_metadata.json} that documents the variables, units, and coverage. The imagery records are raster files, each accompanied by a metadata JSON file of the same basename, with georeferenced tiles additionally carrying their coordinate reference system and bounds for reading with geospatial raster libraries. The geospatial records are provided both as GeoJSON FeatureCollections, which can be read with a geospatial vector library, and as node and relationship CSV tables under a \texttt{neo4j\_import} directory that can be loaded directly into a graph database. A minimal loading example for each modality is shown in Listing~\ref{lst:loading}.

\begin{lstlisting}[language=Python, caption={Minimal per-modality data loading (file paths are illustrative).}, label={lst:loading}]
import json
import pandas as pd
import geopandas as gpd
from PIL import Image

# Textual: one JSON object per line
docs = pd.read_json("textual/wiki.jsonl", lines=True)

# Numerical: CSV table plus co-located metadata JSON
series = pd.read_csv("numerical/entsoe/entsoe_load_DE_2023.csv")
with open("numerical/entsoe/entsoe_load_DE_2023_metadata.json") as f:
    series_meta = json.load(f)

# Imagery: raster file plus same-basename metadata JSON
image = Image.open("imagery/copernicus/copernicus_DE_0001.jpg")
with open("imagery/copernicus/copernicus_DE_0001.json") as f:
    image_meta = json.load(f)

# Geospatial: GeoJSON vectors, and graph-ready CSVs for a graph database
assets = gpd.read_file("geospatial/osm/geojson/DE_power_plant.geojson")
nodes  = pd.read_csv("geospatial/osm/neo4j_import/power_plant_nodes.csv")
\end{lstlisting}

The dataset supports several typical use cases. First, the textual corpus can be used for continual pre-training or domain-adaptive fine-tuning of large language models on energy-specific language. Second, the dataset can power retrieval-augmented generation, since the four modalities are indexed into the OpenSearch, Milvus, and Neo4j back-ends provided with the released code \cite{maienergy2024vectordatabases}, and a hybrid retriever combines semantic, lexical, image, and graph search into a single ranking, as described in the cross-modal linkage subsection. Third, because the images are embedded in a shared text-image space, the imagery modality supports text-to-image retrieval, for example finding satellite tiles or energy labels relevant to a textual query. Fourth, the geospatial and relational modality supports knowledge-graph analytics, such as querying infrastructure assets by country, tracing cross-border interconnections, or exploring research-project collaborations. Fifth, the shared entity keys of country, bidding zone, and geolocation allow cross-modal analyses that join, for example, electricity-system time series, the power plants located in the same country, and the policy documents that discuss them. Finally, the numerical modality alone supports conventional energy-system tasks such as load and price analysis or time-series forecasting.

To make this use concrete, the released data and back-ends support a retrieval-augmented generation workflow. A natural-language energy question is issued to the hybrid retriever described in the cross-modal linkage subsection, which returns evidence from several modalities at once, and the retrieved evidence is supplied as context to a large language model that produces a grounded answer. For the example question \emph{``What is the role of renewable energy in Austria's electricity system?''}, the top item returned from each contributing store is summarised in Table~\ref{tab:rag-example}, and the answer composed by the model from this evidence was \emph{``Austria's electricity system is predominantly renewable and led by hydropower, with wind and solar contributing a growing share, consistent with European renewable energy targets.''}

\begin{table}[ht]
\centering
\caption{Evidence retrieved across modalities by the hybrid retriever for the example question, with the top item from each contributing store.}
\label{tab:rag-example}
\begin{tabularx}{\textwidth}{
  >{\RaggedRight\arraybackslash}p{2.6cm}
  >{\RaggedRight\arraybackslash}p{2.3cm}
  >{\RaggedRight\arraybackslash}X
}
\toprule
\textbf{Modality} & \textbf{Store} & \textbf{Top retrieved item} \\
\midrule
Textual & OpenSearch & Wikipedia article ``Renewable energy in Austria'', describing the hydropower-dominated electricity mix (\url{https://en.wikipedia.org/wiki/Renewable_energy_in_Austria}). \\
Numerical & OpenSearch & Eurostat renewable energy share series for Austria (AT), share of renewable sources in gross electricity consumption. \\
Imagery & Milvus & Copernicus land-cover tile covering eastern Austria, from the Copernicus Land Monitoring Service. \\
Geospatial / graph & Neo4j & \texttt{PowerPlant} node ``Freudenau'' (run-of-river hydropower) with a \texttt{LOCATED\_IN} edge to the Austria country node. \\
\bottomrule
\end{tabularx}
\end{table}

A limitation of the current release concerns its geographic scope. Because the dataset was assembled around the European energy system and draws heavily on European institutional open-data sources, its coverage is predominantly European. The electricity-system, statistical, building-stock, and weather records, the Copernicus and EPREL imagery, the European transmission-grid and TSO-network data, and the EU governmental and CORDIS sources are all centred on the European Union, whereas the encyclopedic, news, scientific, and contextual or façade imagery sources are global in scope but topically focused on European energy themes, and only a small share of the data extends clearly beyond Europe, most notably the aerial imagery, which covers selected United States cities in addition to European ones. As a result, large language model and retrieval-augmented generation applications built on the dataset as released are expected to be most reliable for European contexts, regulatory regimes, and reporting conventions, and their outputs may be less representative for regions whose energy systems, regulations, and data conventions differ. Users targeting non-European settings should take this imbalance into account when interpreting results or evaluating models. Because the schema and the retrieval pipelines are not specific to Europe, coverage can be broadened by re-running the globally scoped pipelines for additional regions and by substituting the region-specific institutional sources with their equivalents for the target region, as described next.

The dataset is also designed to be extended. Users can re-run the modality-specific retrieval pipelines with different seed topics, keywords, queries, regions, or time ranges to broaden the corpus, or integrate proprietary data into the same schema and back-ends, using the released retrieval code \cite{maienergy2024dataretrieval}.

\section*{Data Availability}

The full dataset release is archived on Zenodo \cite{mylonas2025maienergy}. The harmonised compilation, the curation and processing, and the metadata produced in this work are released under the Creative Commons Attribution 4.0 International (CC BY 4.0) license. Each constituent source retains its own upstream license, and the applicable licenses, together with the required attributions, are documented per source in the retrieval repositories \cite{maienergy2024dataretrieval} and consolidated in the dataset record. As exceptions to the overall CC BY 4.0 release, files derived from OpenStreetMap and from GridKit are made available under the Open Database License (ODbL) 1.0, and text derived from Wikipedia is made available under CC BY-SA 4.0.

\section*{Code Availability}

The complete data retrieval and preparation pipelines for all four modalities are openly available on GitLab \cite{maienergy2024dataretrieval}, organised as one repository per modality within the \texttt{maienergy-data-retrieval} group:
\begin{itemize}
    \item Textual articles retrieval: \url{https://gitlab.com/maienergy-data-retrieval/articles-retrieval}
    \item Numerical retrieval: \url{https://gitlab.com/maienergy-data-retrieval/numerical-retrieval}
    \item Geospatial retrieval: \url{https://gitlab.com/maienergy-data-retrieval/geospatial-retrieval}
    \item Imagery retrieval: \url{https://gitlab.com/maienergy-data-retrieval/images-retrieval}
\end{itemize}

\noindent In addition, the database back-ends used to index the corpus and operationalise it for retrieval are provided in the \texttt{maienergy-vector-databases} group \cite{maienergy2024vectordatabases}:
\begin{itemize}
    \item Milvus, a dense-vector database for semantic retrieval: \url{https://gitlab.com/maienergy-vector-databases/milvus-setup}
    \item Neo4j, a graph database for cross-modal node and relationship ingestion: \url{https://gitlab.com/maienergy-vector-databases/neo4j-setup}
    \item OpenSearch, for hybrid lexical and vector search: \url{https://gitlab.com/maienergy-vector-databases/opensearch-setup}
\end{itemize}

\bibliographystyle{elsarticle-num} 
\bibliography{references}

@article{majumder2024exploring,
  title={Exploring the capabilities and limitations of large language models in the electric energy sector},
  author={Majumder, Subir and Dong, Lin and Doudi, Fatemeh and Cai, Yuting and Tian, Chao and Kalathil, Dileep and Ding, Kevin and Thatte, Anupam A and Li, Na and Xie, Le},
  journal={Joule},
  volume={8},
  number={6},
  pages={1544--1549},
  year={2024},
  publisher={Elsevier}
}

@article{hirth2018entso,
  title={{The ENTSO-E Transparency Platform--A review of Europe’s most ambitious electricity data platform}},
  author={Hirth, Lion and M{\"u}hlenpfordt, Jonathan and Bulkeley, Marisa},
  journal={Applied energy},
  volume={225},
  pages={1054--1067},
  year={2018},
  publisher={Elsevier}
}

@misc{ENTSOE,
  author = {{ENTSO-E}},
  title = {{ENTSO-E Transparency Platform}},
  howpublished = {\url{https://transparency.entsoe.eu}},
}

@misc{Eurostat,
  author = {{Eurostat}},
  title = {Eurostat Statistics},
  howpublished = {\url{https://ec.europa.eu/eurostat/data/database}}
}

@misc{copernicus_discomap,
  author       = {{European Environment Agency (EEA)}},
  title        = {{European Union's Copernicus Land Monitoring Service imagery (Discomap)}},
  howpublished = {\url{https://image.discomap.eea.europa.eu/}}
}

@article{haklay2008openstreetmap,
  title={Openstreetmap: User-generated street maps},
  author={Haklay, Mordechai and Weber, Patrick},
  journal={IEEE Pervasive computing},
  volume={7},
  number={4},
  pages={12--18},
  year={2008},
  publisher={Ieee}
}

@article{lewis2020retrieval,
  title={Retrieval-augmented generation for knowledge-intensive nlp tasks},
  author={Lewis, Patrick and Perez, Ethan and Piktus, Aleksandra and Petroni, Fabio and Karpukhin, Vladimir and Goyal, Naman and K{\"u}ttler, Heinrich and Lewis, Mike and Yih, Wen-tau and Rockt{\"a}schel, Tim and others},
  journal={Advances in neural information processing systems},
  volume={33},
  pages={9459--9474},
  year={2020}
}

@article{wilkinson2016fair,
  title={{The FAIR Guiding Principles for scientific data management and stewardship}},
  author={Wilkinson, Mark D and Dumontier, Michel and Aalbersberg, IJsbrand Jan and Appleton, Gabrielle and Axton, Myles and Baak, Arie and Blomberg, Niklas and Boiten, Jan-Willem and da Silva Santos, Luiz Bonino and Bourne, Philip E and others},
  journal={Scientific data},
  volume={3},
  number={1},
  pages={1--9},
  year={2016},
  publisher={Nature Publishing Group}
}

@article{pfenninger2017importance,
  title={The importance of open data and software: Is energy research lagging behind?},
  author={Pfenninger, Stefan and DeCarolis, Joseph and Hirth, Lion and Quoilin, Sylvain and Staffell, Iain},
  journal={Energy Policy},
  volume={101},
  pages={211--215},
  year={2017},
  publisher={Elsevier}
}

@dataset{mylonas2025maienergy,
  author = {Costas Mylonas and Magda Foti},
  title = {{mAIEnergy Dataset: Multimodal Energy Data for LLM Fine‑Tuning and Retrieval‑Augmented Generation (RAG)}},
  year = {2025},
  publisher    = {Zenodo},
  howpublished = {\url{https://doi.org/10.5281/zenodo.16401633}}
}

@misc{wikipedia,
  author = {{Wikipedia contributors}},
  title = {{Wikipedia: The Free Encyclopedia}},
  publisher = {Wikimedia Foundation},
  howpublished = {\url{https://www.wikipedia.org}}
}

@misc{gnews,
  author = {{Google News API (GNews)}},
  title = {{GNews API}},
  howpublished = {\url{https://gnews.io}}
}

@misc{arxiv,
  author = {{arXiv.org e-Print archive}},
  title = {{arXiv}},
  publisher = {Cornell University},
  howpublished = {\url{https://arxiv.org}}
}

@misc{eu_dg_energy,
  author = {{European Commission -- Directorate-General for Energy}},
  title = {{EU DG Energy Portal}},
  howpublished = {\url{https://energy.ec.europa.eu/index_en}}
}

@misc{acer,
  author = {{Agency for the Cooperation of Energy Regulators (ACER)}},
  title = {{ACER Website}},
  howpublished = {\url{https://acer.europa.eu/}}
}

@misc{inria_dataset,
  author = {{INRIA}},
  title = {{INRIA Aerial Image Labeling Dataset}},
  howpublished = {\url{https://project.inria.fr/aerialimagelabeling}}
}

@misc{eprel,
  author = {{European Commission}},
  title = {{European Product Registry for Energy Labelling (EPREL)}},
  howpublished = {\url{https://eprel.ec.europa.eu/}}
}

@misc{irf_dataset,
  author = {S. Rahmani and I. Benenson},
  title = {{IRF: Irregular Façade Dataset}},
  publisher = {Kaggle},
  howpublished = {\url{https://www.kaggle.com/datasets/saeedrahmani/irregular-facades}}
}

@misc{wikimedia_commons,
  author = {{Wikimedia Commons}},
  title = {{Wikimedia Commons}},
  publisher = {Wikimedia Foundation},
  howpublished = {\url{https://commons.wikimedia.org}}
}

@misc{eu_building_stock,
  author = {{European Commission}},
  title = {{EU Building Stock Observatory}},
  howpublished = {\url{https://ec.europa.eu/energy/eu-buildings-database_en}}
}

@misc{open_meteo,
  author = {{Open-Meteo}},
  title = {{Open-Meteo Historical Weather Data}},
  howpublished = {\url{https://open-meteo.com}}
}

@misc{gridkit,
  author = {Bart Wiegmans},
  title = {{GridKit European Transmission Grid}},
  howpublished = {\url{https://zenodo.org/records/47317}}
}

@misc{wri_powerplants,
  author = {{World Resources Institute}},
  title = {{Global Power Plant Database}},
  howpublished = {\url{https://datasets.wri.org/dataset/globalpowerplantdatabase}}
}

@misc{cordis,
  author = {{CORDIS}},
  title = {{CORDIS EU Projects Database}},
  publisher = {European Commission},
  howpublished = {\url{https://cordis.europa.eu}}
}

@article{taylor2022galactica,
  title={Galactica: A large language model for science},
  author={Taylor, Ross and Kardas, Marcin and Cucurull, Guillem and Scialom, Thomas and Hartshorn, Anthony and Saravia, Elvis and Poulton, Andrew and Kerkez, Viktor and Stojnic, Robert},
  journal={arXiv preprint arXiv:2211.09085},
  year={2022}
}

@article{antonesi2025transformers,
  title={{From Transformers to Large Language Models: A systematic review of AI applications in the energy sector towards Agentic Digital Twins}},
  author={Antonesi, Gabriel and Cioara, Tudor and Anghel, Ionut and Michalakopoulos, Vasilis and Sarmas, Elissaios and Toderean, Liana},
  journal={arXiv preprint arXiv:2506.06359},
  year={2025}
}

@inproceedings{jatowt2025flexidigital,
  title={{FlexiDigital}: A Comprehensive Approach to Energy Flexibility Services using Digital Twins and Large Language Models},
  author={Jatowt, Adam and Ristov, Sashko and Gritsch, Philipp and Brandacher, Simon and Rosengren, Peter and Valerio, Danilo and Luo, Fengji},
  booktitle={Proceedings of the 16th ACM International Conference on Future and Sustainable Energy Systems},
  pages={638--643},
  year={2025}
}

@inproceedings{lin2024open,
  title={An open and large-scale dataset for multi-modal climate change-aware crop yield predictions},
  author={Lin, Fudong and Guillot, Kaleb and Crawford, Summer and Zhang, Yihe and Yuan, Xu and Tzeng, Nian-Feng},
  booktitle={Proceedings of the 30th ACM SIGKDD Conference on Knowledge Discovery and Data Mining (KDD)},
  pages={5375--5386},
  year={2024}
}

@article{zhang2024global,
  title={A global multimodal flood event dataset with heterogeneous text and multi-source remote sensing images},
  author={Zhang, Zhixin and Ma, Yan and Liu, Peng},
  journal={Big Earth Data},
  pages={1--27},
  year={2024},
  publisher={Taylor \& Francis}
}

@misc{maienergy2024dataretrieval,
  author = {Costas Mylonas},
  title = {{mAiEnergy Data Retrieval}},
  year = {2024},
  howpublished = {\url{https://gitlab.com/maienergy-data-retrieval}}
}

@misc{maienergy2024vectordatabases,
  author       = {Costas Mylonas},
  title        = {{mAIEnergy Vector Databases}},
  year         = {2024},
  howpublished = {\url{https://gitlab.com/maienergy-vector-databases}}
}

@article{silva2020benchmark,
  title={Benchmark maps of 33 years of secondary forest age for Brazil},
  author={Silva Junior, Celso HL and Heinrich, Viola HA and Freire, Ana TG and Broggio, Igor S and Rosan, Thais M and Doblas, Juan and Anderson, Liana O and Rousseau, Guillaume X and Shimabukuro, Yosio E and Silva, Carlos A and others},
  journal={Scientific data},
  volume={7},
  number={1},
  pages={269},
  year={2020},
  publisher={Nature Publishing Group UK London}
}

@article{milojevic2023eubucco,
  title={{EUBUCCO} v0.1: European building stock characteristics in a common and open database for 200+ million individual buildings},
  author={Milojevic-Dupont, Nikola and Wagner, Felix and Nachtigall, Florian and Hu, Jiawei and Br{\"u}ser, Geza Boi and Zumwald, Marius and Biljecki, Filip and Heeren, Niko and Kaack, Lynn H and Pichler, Peter-Paul and others},
  journal={Scientific data},
  volume={10},
  number={1},
  pages={147},
  year={2023},
  publisher={Nature Publishing Group UK London}
}

@article{mitchell2022measuring,
  title={Measuring data},
  author={Mitchell, Margaret and Luccioni, Alexandra Sasha and Lambert, Nathan and Gerchick, Marissa and McMillan-Major, Angelina and Ozoani, Ezinwanne and Rajani, Nazneen and Thrush, Tristan and Jernite, Yacine and Kiela, Douwe},
  journal={arXiv preprint arXiv:2212.05129},
  year={2022}
}

@article{schuhmann2022laion,
  title={Laion-5b: An open large-scale dataset for training next generation image-text models},
  author={Schuhmann, Christoph and Beaumont, Romain and Vencu, Richard and Gordon, Cade and Wightman, Ross and Cherti, Mehdi and Coombes, Theo and Katta, Aarush and Mullis, Clayton and Wortsman, Mitchell and others},
  journal={Advances in neural information processing systems},
  volume={35},
  pages={25278--25294},
  year={2022}
}

@article{fischer2023approaches,
  title={Approaches and tools for user-driven provenance and data quality information in spatial data infrastructures},
  author={Fischer, Julia and Egli, Lukas and Groth, Juliane and Barrasso, Caterina and Ehrmann, Steffen and Figgemeier, Heiko and Henzen, Christin and Meyer, Carsten and M{\"u}ller-Pfefferkorn, Ralph and R{\"u}mmler, Arne and others},
  journal={International Journal of Digital Earth},
  volume={16},
  number={1},
  pages={1510--1529},
  year={2023},
  publisher={Taylor \& Francis}
}

@misc{ISO19157-2013,
  title        = {{ISO 19157:2013 -- Geographic information – Data quality}},
  institution  = {{International Organization for Standardization (ISO)}},
  year         = {2013},
  month        = dec,
  howpublished = {\url{https://www.iso.org/standard/32575.html}}
}

@article{emami2023buildingsbench,
  title={Buildingsbench: A large-scale dataset of 900k buildings and benchmark for short-term load forecasting},
  author={Emami, Patrick and Sahu, Abhijeet and Graf, Peter},
  journal={Advances in Neural Information Processing Systems},
  volume={36},
  pages={19823--19857},
  year={2023}
}

@article{wiese2019open,
  title={Open Power System Data--Frictionless data for electricity system modelling},
  author={Wiese, Frauke and Schlecht, Ingmar and Bunke, Wolf-Dieter and Gerbaulet, Clemens and Hirth, Lion and Jahn, Martin and Kunz, Friedrich and Lorenz, Casimir and M{\"u}hlenpfordt, Jonathan and Reimann, Juliane and others},
  journal={Applied Energy},
  volume={236},
  pages={401--409},
  year={2019},
  publisher={Elsevier}
}

@article{zhu2022meter,
  title={Meter-ml: A multi-sensor earth observation benchmark for automated methane source mapping},
  author={Zhu, Bryan and Lui, Nicholas and Irvin, Jeremy and Le, Jimmy and Tadwalkar, Sahil and Wang, Chenghao and Ouyang, Zutao and Liu, Frankie Y and Ng, Andrew Y and Jackson, Robert B},
  journal={arXiv preprint arXiv:2207.11166},
  year={2022}
}

@article{schmitt2019sen12ms,
  title={SEN12MS--A curated dataset of georeferenced multi-spectral sentinel-1/2 imagery for deep learning and data fusion},
  author={Schmitt, Michael and Hughes, Lloyd Haydn and Qiu, Chunping and Zhu, Xiao Xiang},
  journal={arXiv preprint arXiv:1906.07789},
  year={2019}
}

@article{webersinke2021climatebert,
  title={Climatebert: A pretrained language model for climate-related text},
  author={Webersinke, Nicolas and Kraus, Mathias and Bingler, Julia Anna and Leippold, Markus},
  journal={arXiv preprint arXiv:2110.12010},
  year={2021}
}

@article{chebbi2026towards,
  title={Towards EnergyGPT: A Large Language Model Specialized for the Energy Sector},
  author={Chebbi, Amal and Kolade, Babajide},
  journal={IEEE Access},
  year={2026},
  publisher={IEEE}
}

@article{zhou2024elecbench,
  title={Elecbench: a power dispatch evaluation benchmark for large language models},
  author={Zhou, Xiyuan and Zhao, Huan and Cheng, Yuheng and Cao, Yuji and Liang, Gaoqi and Liu, Guolong and Liu, Wenxuan and Xu, Yan and Zhao, Junhua},
  journal={arXiv preprint arXiv:2407.05365},
  year={2024}
}

\section*{Author Contributions}

\noindent Conceptualization: C.M., M.F.; Methodology: C.M.; Software: C.M.; Validation: C.M.; Formal analysis: C.M.; Investigation: C.M.; Data curation: C.M.; Writing – original draft preparation: C.M.; Writing – review and editing: C.M., M.F.; Visualization: C.M.; Supervision: M.F.

\section*{Competing Interests}

\noindent The authors declare no competing interests.

\section*{Funding}

\noindent This work was supported by European Union’s funded project DIGITISE under grant agreement No 01160671 and the FFplus project, which is funded by the European High-Performance Computing Joint Undertaking (JU) under grant agreement No 101163317.

\end{document}